\documentclass[aps,print,prb,twocolumn,superscriptaddress]{revtex4-2}

\usepackage[hidelinks]{hyperref}
\hypersetup{
	colorlinks,
	linkcolor={red},
	citecolor={blue},
	urlcolor={blue}
}
\usepackage{physics}
\usepackage{balance}
\usepackage{amssymb}
\usepackage{suffix}
\usepackage{mathtools}
\usepackage[utf8]{inputenc}
\usepackage{booktabs}
\usepackage{cases}
\usepackage[multiple]{footmisc}
\usepackage{dcolumn}
\usepackage{color,soul}
\usepackage{rotating}
\usepackage{perpage}
\usepackage{siunitx}
\usepackage{xcolor}
\usepackage{soul}
\usepackage{amsmath}
\usepackage{tikz}
\usepackage[T1]{fontenc}
\usepackage{etoolbox}
\usepackage{graphics}
\usepackage{siunitx}
\usepackage{float}	
\usepackage{collref}
\usepackage{multirow}
\usepackage{mathtools}
\usepackage{bm}
\usepackage{url}

\usepackage{tikz}
\usepackage{tikz-3dplot}
\usepackage{accents}

\makeatletter
\newcommand{\doublewidetilde}[1]{{%
		\mathpalette\double@widetilde{#1}%
}}
\newcommand{\double@widetilde}[2]{%
	\sbox\z@{$\m@th#1\widetilde{#2}$}%
	\ht\z@=.9\ht\z@
	\widetilde{\box\z@}%
}
\makeatother

\usepackage{etoolbox,lipsum}

\DeclarePairedDelimiterX\MeijerM[3]{\lparen}{\rparen}%
{\begin{smallmatrix}#1 \\ #2\end{smallmatrix}\delimsize\vert\,#3}

\newcommand\MeijerG[8][]{%
	\mathbb{G}^{\,#2,#3}_{#4,#5}\MeijerM[#1]{#6}{#7}{#8}}

\WithSuffix\newcommand\MeijerG*[7]{%
	\mathbb{G}^{\,#1,#2}_{#3,#4}\MeijerM*{#5}{#6}{#7}}

\begin{document} 
	\title{Edelstein effect in optically driven monolayer jacutingaite Pt$_2$HgSe$_3$}
	
	\author{Nguyen Q. Bau}
	\address{Department of Theoretical Physics, Faculty of Physics,
		University of Science, Vietnam National University, Hanoi,
		334 Nguyen Trai, Thanh Xuan, Hanoi 100000, Viet Nam}
		\author{Ta T. Tho}
		\address{Department of Physics, Faculty of Mechanical Engineering,
			Hanoi University of Civil Engineering, Hanoi, Vietnam}
		\author{Le T. T. Phuong}
		\address{Faculty of of Physics, University of Education, Hue University, Hue 530000, Vietnam}
	\author{Bui D. Hoi}
	\email{Corresponding author: buidinhhoi@hueuni.edu.vn}
	\address{Faculty of of Physics, University of Education, Hue University, Hue 530000, Vietnam}

	\date{\today}
	\begin{abstract}
The optical control of spin- and valley-selective gapless states in two-dimensional materials presents new opportunities for next-generation spintronic and valleytronic technologies. In this work, we study monolayer jacutingaite (Pt$_2$HgSe$_3$), a quantum spin Hall insulator with strong intrinsic spin-orbit coupling, under irradiation by circularly polarized light. The light-induced Floquet engineering gives rise to tunable topological phases, including transitions to spin- and valley-polarized semimetallic states. To probe these topological transitions, we employ the spin and orbital Edelstein effects—non-equilibrium responses arising from spin-orbit interactions in systems lacking inversion symmetry—without resorting to topological invariants such as Chern numbers. We identify universal signatures of the phase transitions encoded in the Edelstein response: a pronounced discontinuity in the spin Edelstein conductivity and a vanishing orbital Edelstein susceptibility mark the onset of the semimetallic regime. Furthermore, we investigate how the growth and suppression of the spin Edelstein responses across the topological phase transition depend on the interband scattering time. These findings establish the Edelstein effect as a sensitive and experimentally accessible probe of light-induced topological transitions in quantum materials.
	\end{abstract}
	
	\maketitle
	{\allowdisplaybreaks
		
		\section{Introduction}
		
		The exploration of topological phases in two-dimensional (2D) materials has garnered growing attention due to their potential applications in spintronics and quantum information technologies~\cite{hasan2010colloquium,qi2011topological}. A key feature of many such systems is the interplay between symmetry breaking and spin-orbit coupling (SOC), which can give rise to a variety of exotic phases, including quantum spin Hall, quantum anomalous Hall, and quantum valley Hall states~\cite{manzeli20172d, xu2014spin}. Among the diverse family of 2D materials, monolayer jacutingaite (Pt$_2$HgSe$_3$) stands out due to its large intrinsic SOC and favorable electronic structure~\cite{PhysRevLett.120.117701,Marrazzo2019,PhysRevResearch.2.012063,PhysRevLett.124.106402,PhysRevB.100.235101}. This compound has been theoretically predicted and experimentally supported as a quantum spin Hall insulator with a substantial band gap and topologically protected edge modes, rendering it a strong candidate for energy-efficient spintronic applications~\cite{facio2019dual,shah2024topological}.
		
		Recent advances have shown that jacutingaite’s topological character is highly sensitive to external perturbations. Static electric fields and circularly polarized light can drive the system through a sequence of quantum phase transitions, modulating its topological invariants and inducing transitions among various phases~\cite{PhysRevB.111.014440,shah2024topological,vargiamidis2022,PhysRevResearch.5.043263}. These light-induced transitions are rooted in the interplay between intrinsic SOC and Floquet engineering, which reshapes the band structure and Berry curvature landscape. Such tunability establishes jacutingaite as a versatile platform for realizing dynamic topological control in next-generation optoelectronic and valleytronic devices.
		
		In parallel, the Edelstein effect—encompassing both spin and orbital variants—has emerged as a hallmark of spin-charge interconversion in non-centrosymmetric systems with strong SOC. It describes the generation of non-equilibrium spin or orbital polarizations in response to an applied electric field~\cite{edelstein1990spin,manchon2015new,lesne2016highly}. While traditionally studied in Rashba interfaces, topological insulators, and heavy-metal/ferromagnet bilayers, yet, a comprehensive understanding of how the Edelstein response encodes information about topological transitions in light-driven 2D systems remains elusive.
		
		Motivated by these developments, we investigate whether the Edelstein effect—particularly its spin and orbital components—can serve as a probe of photoinduced topological phase transitions in monolayer jacutingaite. Given the sensitivity of Edelstein conductivities to changes in band topology, analyzing their behavior under varying light irradiation offers a powerful, symmetry-sensitive diagnostic. This study aims to establish a connection between optically tunable topological phases and the Edelstein response in monolayer jacutingaite. Through theoretical modeling and numerical simulations, we systematically characterize the evolution of spin and orbital Edelstein conductivities across photoinduced phase boundaries. Our findings highlight the diagnostic potential of the Edelstein effect for detecting and characterizing topological transitions, offering new insights into the control of quantum phases in 2D materials and paving the way for advanced optospintronic functionalities.
		
		The remainder of this paper is organized as follows. In Sec. \ref{s2}, we introduce the effective low-energy model describing jacutingaite under static and dynamic electric fields, and outline the formalism used to evaluate Edelstein conductivities. Section \ref{s3} presents numerical results for spin and orbital Edelstein responses, emphasizing their connection to the underlying band topology. Finally, Sec. \ref{s4} concludes with a summary and an outlook on future directions.
		
		\section{Theoretical Framework}\label{s2}
		
		\subsection{Hamiltonian model}
		We begin with the light-dressed low-energy Dirac Hamiltonian describing a two-band model for electrons in a honeycomb lattice structure of monolayer jacutingaite Pt$_2$HgSe$_3$, incorporating spin and valley degrees of freedom. Each valley $\zeta = \pm 1$ corresponds to $K$ and $K'$ points in the Brillouin zone, and each spin $s = \pm 1$ labels the two spin orientations. 	It is worth noting that additional external exchange fields can be incorporated into the model, and the same set of calculations can be repeated. However, the main conclusion of the paper remains unchanged. To avoid redundancy, we focus exclusively on the effects of light in this work. For discussions involving other types of external fields, see Refs.~\cite{PhysRevB.111.014440,vargiamidis2022,shah2024topological,PhysRevResearch.5.043263}. The Hamiltonian reads:
		\begin{align}\label{eq_1}
			H_{\zeta,s}(\vec{k}) = v_F (\zeta k_x \sigma_x + k_y \sigma_y) + \Delta_{\zeta,s} \sigma_z\, ,
		\end{align}where $v_F \approx 3\times10^5$ m/s is the Fermi velocity, $\sigma_i$ are the Pauli matrices acting on the pseudospin (sublattice) space, and $\Delta_{\zeta,s}$ is a mass term defined by:
		\begin{align}\label{eq_2}
			\Delta_{\zeta,s} = \zeta s \Delta_{so} + \zeta \Delta_d\, .
		\end{align}This term captures SOC ($\Delta_{so}$) and a possible valley-contrasting light-induced potential ($\Delta_d$). We aim to study spin and orbital Edelstein effects within the linear response regime. See Appendix~\ref{ap1} for derivation of the light-induced potential.
		
		The Hamiltonian can be recast as $H_{\zeta,s}(\vec{k}) = \vec{d}(\vec{k}) \cdot \vec{\sigma}$ using
		\begin{subequations}
			\begin{align}
				\vec{d}(\vec{k}) = {} &(v_F \zeta k_x, v_F k_y, \Delta_{\zeta,s})\, ,\\ \quad d(\vec{k}) = {} & |\vec{d}(\vec{k})| = \sqrt{v_F^2 k^2 + \Delta_{\zeta,s}^2}\, .
			\end{align}
		\end{subequations}
		The eigenenergies are:
		\begin{align}
			\varepsilon_{n\vec{k}} = n d(\vec{k}), \quad n = \pm 1\,.
		\end{align}
		Also, the corresponding eigenstates, expressed as two-component spinors in the pseudospin basis, are:
		\begin{subequations}
			\begin{align}
				|u_{+\vec{k}}\rangle ={} & \begin{pmatrix} \cos(\theta_k/2) \\\\ \sin(\theta_k/2)e^{i\phi_k} \end{pmatrix}\, ,\\
				|u_{-\vec{k}}\rangle = {} &\begin{pmatrix} \sin(\theta_k/2) \\\\ -\cos(\theta_k/2)e^{i\phi_k} \end{pmatrix},
			\end{align}
		\end{subequations}with:
		\begin{align}
			\cos \theta_k = \frac{\Delta_{\zeta,s}}{d}, \quad \sin \theta_k = \frac{v_F k}{d}, \quad \tan \phi_k = \frac{k_y}{\zeta k_x}.
		\end{align}
		\begin{figure}[t]
			\centering
			\includegraphics[width=1\linewidth]{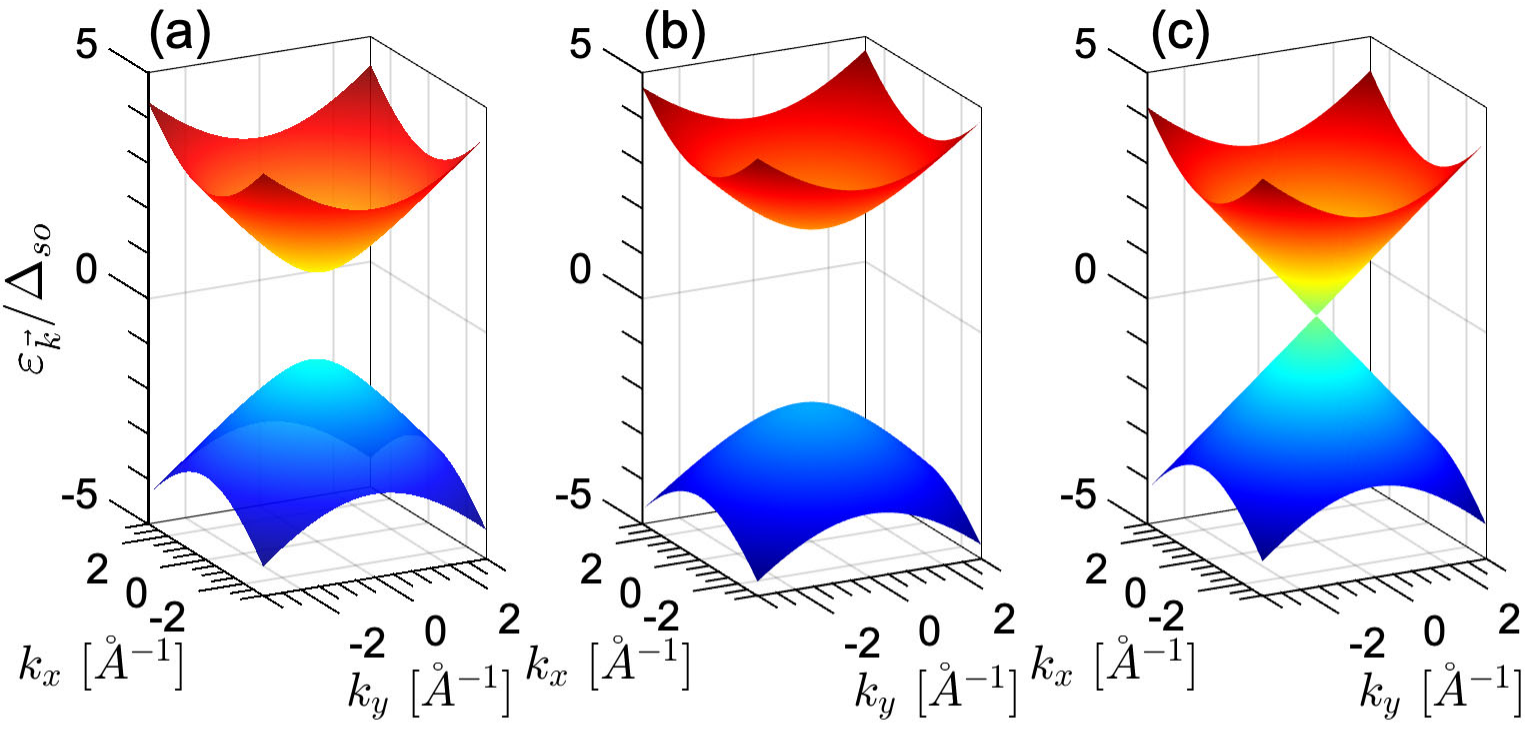}
			\caption{Electronic band structure of monolayer jacutingaite Pt\(_2\)HgSe\(_3\) near the valley \(K\): (a) both spin-up and spin-down states in the absence of light, (b) spin-up state under illumination with \(\Delta_d/\Delta_{\text{so}} = 1\), and (c) spin-down state under the same light-induced condition. In the absence of light, the mass term in Eq.~\eqref{eq_2} is given by \(\Delta_{+1,s}/ \Delta_{\text{so}} = s\), resulting in identical band gaps for both spin states with opposite signs. Upon light irradiation at \(\Delta_d/\Delta_{\text{so}} = 1\), the mass term becomes \(\Delta_{+1,s} /\Delta_{\text{so}} = s + 1\), which vanishes for spin-down (\(s = -1\)), leading to a gapless, semimetallic phase. Note that the situation is reversed when \(\Delta_d/\Delta_{\text{so}} = -1\), in which case the mass term vanishes for spin-up instead. A similar gap closing occurs for spin-up in the opposite valley \(K'\) (not shown).} 
			\label{f1n}
		\end{figure}
		
		Figure~\ref{f1n} illustrates the light-tunable electronic band structure of monolayer jacutingaite Pt$_2$HgSe$_3$ near the valley $K$, showcasing how spin-valley selective gap closings can be induced optically. Panel (a) shows the pristine band structure in the absence of light, where both spin-up and spin-down states exhibit symmetric gapped Dirac dispersions. These gaps originate from spin-orbit coupling, yielding mass terms of opposite signs for each spin species, $\Delta_{+1,s}/\Delta_{\rm so} = s$. As a result, the system supports a quantum spin Hall insulating phase, characterized by equal gap magnitudes and opposite Berry curvatures for opposite spins. Panels (b) and (c) depict the system under illumination with $\Delta_d/\Delta_{\rm so} = 1$. For the spin-up state in panel (b), the mass term becomes $\Delta_{+1,\uparrow}/\Delta_{\rm so} = +2$, enlarging the gap and further localizing the Dirac fermions. In contrast, for the spin-down state in panel (c), the mass term becomes $\Delta_{+1,\downarrow}/\Delta_{\rm so} = 0$, resulting in a completely closed gap and a linear, massless Dirac cone. This corresponds to a light-induced semimetallic phase for spin-down electrons in valley $K$, highlighting the potential to achieve spin-polarized gapless transport through optical control. Although not shown here, the same analysis applies to the opposite valley $K'$, where a similar gap closing occurs for the spin-up state at the same light intensity. This reflects the valley-contrasting nature of the light-induced potential, enabling full spin-valley locking. 
		
		\subsection{Spin and orbital Edelstein effects}
		The Edelstein effect refers to the generation of a non-equilibrium spin or orbital polarization in response to an applied electric field in systems with strong SOC and broken inversion symmetry. In 2D materials, this effect arises from the spin-momentum locking encoded in the band structure, and it manifests in both spin and orbital channels. The spin Edelstein effect is characterized by a current-induced spin polarization, whereas the orbital Edelstein effect involves a current-induced orbital magnetic moment~\cite{edelstein1990spin}. 		
		
		Within linear response theory, the induced expectation value of the spin (or orbital angular momentum) operator $\hat{O}_\alpha$ in direction $\alpha$ due to an electric field \( E_\beta \) along direction $\beta$ is given by
			\begin{align}
			\langle \hat{O}^i \rangle = \sum_\beta \chi_{i j}  E^j\,,
		\end{align}where \( \chi_{i j} \) is the Edelstein conductivity tensor. For the spin Edelstein effect, \( \hat{O}^i = \hat{S}^i \), while for the orbital Edelstein effect, \( \hat{O}^i = \hat{L}^i\). The tensor \( \chi_{i j} \) can be calculated using the Kubo formula in the static limit. These tensor elements encode both intraband and interband coherence between states mediated by SOC, making the Edelstein response a sensitive probe of topological and symmetry-breaking phenomena. To compute the spin Edelstein effect and orbital Edelstein effect, we need matrix elements of the velocity operator: $\hat{v}^i = \hbar^{-1}\frac{\partial H}{\partial k_i}$, spin operator: $\hat{S}^i = \frac{\hbar}{2} \sigma_i$, and orbital angular momentum: $\hat{L}^z$, defined via Berry curvature or gauge-invariant band geometry. Within the Kubo formalism, the response function $\chi_{ij}$ (representing spin or orbital susceptibility) is given by~\cite{doi:10.1143/JPSJ.12.570,PhysRevMaterials.5.074407,PhysRevMaterials.6.095001}:\begin{align}
				\chi_{ij} = {} &\frac{ie}{m_e} \int \frac{d^2k}{(2\pi)^2}
				\Bigg[\sum_n \frac{\partial f_{n\vec{k}}}{\partial \varepsilon_{n\vec{k}}} \frac{O^i_{nn,\vec{k}} \,p^j_{nn,\vec{k}} }{i\tau^{-1}_{\text{intra}}}\notag \\ {} &-\sum_{n \neq m} \frac{f_{n\vec{k}} - f_{m\vec{k}}}{\varepsilon_{n\vec{k}}-\varepsilon_{m\vec{k}}} 
			 \frac{O^i_{mn,\vec{k}}\, p^j_{nm,\vec{k}}}{\varepsilon_{m\vec{k}}-\varepsilon_{n\vec{k}} + i\tau^{-1}_{\text{inter}}} 
				 \Bigg]\, ,
			\end{align}where  \( f_{n\vec{k}} \) is the Fermi-Dirac distribution. The first term is the intraband contribution, the second is interband. We define matrix elements:
		\begin{align}
			O^i_{mn,\vec{k}} &= \langle u_{m\vec{k}} | \hat{O}^i | u_{n\vec{k}} \rangle\,, \quad
			p^j_{nm,\vec{k}} &= \langle u_{n\vec{k}} | \hat{v}^j | u_{m\vec{k}} \rangle\,.
		\end{align}At $T \approx 0$, we have $\frac{\partial f_n}{\partial \varepsilon} = -\delta(\varepsilon_n - \mu)$~(with $\mu$ being the chemical potential tuning the Fermi energy), leading to:
		\begin{align}
			\chi_{ij}^{\text{intra}} = - \frac{e\,\tau_{\text{intra}}}{m_e} \int \frac{d^2k}{(2\pi)^2} \sum_n \delta(\varepsilon_n - \mu) O^i_{nn\vec{k}} \,p^j_{nn\vec{k}}\,.
		\end{align}This vanishes if either operator lacks diagonal matrix elements. With  $\hat{O}^i = \frac{\hbar}{2} \sigma_i$, we can compute the spin Edelstein susceptibility for both intraband and interband contributions.
		
		In the case of the orbital Edelstein susceptibility, we consider $\hat{O}^i = L^i_{n\vec{k}}$, which requires the orbital angular momentum, obtained from $\vec{L}_{n\vec{k}} = i \bra{\nabla_{\vec{k}} u_{n\vec{k}}} \times [H_{\zeta,s}(\vec{k}) - \varepsilon_{n\vec{k}}] \ket{\nabla_{\vec{k}} u_{n\vec{k}}}$~\cite{PhysRevB.53.7010,PhysRevB.102.235426}, which corresponds to the internal circulation of the Bloch wave packet~\cite{PhysRevLett.97.026603}. In 2D, the orbital magnetic moment $m^z_{n\vec{k}}$ relates to the Berry curvature $\Omega^z_n(\vec{k})$:
		\begin{subequations}
			\begin{align}
				m^z_{n\vec{k}} = {} &-\frac{e}{\hbar} \, \varepsilon_{n\vec{k}} \, \Omega^z_n(\vec{k})\,,\\
				\Omega^z_n(\vec{k}) = {} &\frac{n}{2} \frac{\vec{d}\cdot\left(\partial_{k_x}\vec{d} \times\partial_{k_y}\vec{d}\,\right)}{|\vec{d}|^3}\, ,
			\end{align}
		\end{subequations}with the dimensionless orbital angular momentum $L^z_{n\vec{k}} = m^z_{n\vec{k}}/\mu_B$ defined as:
		\begin{align}\label{eq_13}
			L^z_{n\vec{k}} = -\frac{e}{\hbar \mu_B} \, \varepsilon_{n\vec{k}}\, \Omega^z_n(\vec{k}) = -n \frac{e \zeta v^2_{F} \Delta_{\zeta,s}}{2 \hbar \mu_{\rm B}\left(v^2_{F}k^2 + \Delta^2_{\zeta,s}\right)}\, ,
		\end{align}where $\mu_B = e\hbar / (2m_e)$ is the Bohr magneton. It is also important to emphasize that, within our model, the interband contributions to the orbital Edelstein susceptibility vanish due to $\langle u_{\pm \vec{k}}|L^z_{\pm \vec{k}} |u_{\mp \vec{k}}\rangle = 0$.
				
		\section{results and discussions}\label{s3}
		Appendix~\ref{ap2} presents the complete set of components required for the calculation of spin and orbital Edelstein susceptibilities. The numerical results are categorized into two sectors corresponding to the vallyes $K$ and $K'$. For simplicity, we set \(\hbar = k_B = m_e = e = 1\) throughout the calculations.\begin{figure*}[t]
			\centering
			\includegraphics[width=1\linewidth]{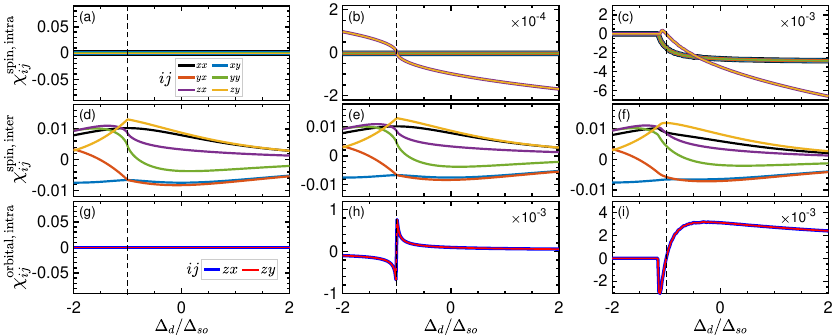}
			\caption{Spin and orbital Edelstein susceptibilities \(\chi_{ij}\)~(in arbitrary units) in monolayer jacutingaite Pt\(_2\)HgSe\(_3\) as functions of the normalized sublattice asymmetry parameter \(\Delta_d/\Delta_{\text{so}}\), shown for three different chemical potentials: \(\mu = 0\) (left), \(\mu = \Delta_{+1,+1}\) (middle), and \(\mu > \Delta_{+1,+1}\) (right). Rows show the intraband spin (top), interband spin (middle), and intraband orbital (bottom) contributions. Intraband spin and orbital responses vanish at charge neutrality due to symmetry, while interband spin contributions remain finite with a discontinuity at \(\Delta_d/\Delta_{\text{so}} = -1\). In the finite chemical potentials, near \(\Delta_d/\Delta_{\text{so}} = -1\), where the mass of one spin-valley flavor vanishes, vanishing features arise in both spin and orbital responses, particularly in the intraband channel. These reflect enhanced Berry curvature and density of states effects at the band inversion point. Intraband responses grow significantly with doping, while interband components are relatively insensitive to \(\mu\).} 
			\label{f2}
		\end{figure*}
		
		To ensure finite intraband contributions to the spin Edelstein effect, the chemical potential is set at zero, at the band edge, $\mu \approx \pm \Delta_{\zeta,s}$, and in the band $\mu > |\Delta_{\zeta,s}|$ at very low temperatures, to capture both electron and hole doping regimes. However, given the electron-hole symmetry of the two-band model, we focus only on the electron-doped regime in the results shown here. 
		
		The lifetime parameters $\tau_{\text{intra}}$ and $\tau_{\text{inter}}$ are first both set to 0.25 eV, which effectively act as decay constants, modeling the interaction of electrons with external environments such as phonon baths. Although the intraband contributions to both the spin and orbital Edelstein susceptibilities scale linearly with $\tau_{\text{intra}}$, we do not focus on its effects in this work. Instead, we will examine the influence of $\tau_{\text{inter}}$ on the interband component of the spin Edelstein susceptibility.
		
		For the $K$ valley ($\zeta = +1$) and spin-up ($s = +1$) states, the mass term takes the form $\Delta_{+1,+1} / \Delta_{\rm so} = 1 + \Delta_d / \Delta_{\rm so}$. When $\Delta_d / \Delta_{\rm so} = -1$, this mass term vanishes, resulting in a band-gap closure and the emergence of a semimetallic phase. This is valid for the $K'$ valley ($\zeta = -1$) and spin-up ($s = +1$) states. According to Eq.~\eqref{eq_2}, spin-up and spin-down states within each valley exhibit symmetric responses, mirrored across the axis defined by $\Delta_d$. For instance, in the same valley ($\zeta = +1$), the spin-down state ($s = -1$) yields a mass term of $\Delta_{+1,-1} / \Delta_{\rm so} = -1 + \Delta_d / \Delta_{\rm so}$, which vanishes when $\Delta_d / \Delta_{\rm so} = 1$. This, again, signals a transition into a semimetallic regime. In the case of valley $K'$ ($\zeta = -1$), the spin-down state ($s = -1$) yields a mass term of $\Delta_{-1,-1} / \Delta_{\rm so} = 1 - \Delta_d / \Delta_{\rm so}$, which vanishes when $\Delta_d / \Delta_{\rm so} = 1$. These observations imply the presence of a spin-polarized semimetal phase at the critical points where the gap closes for only one spin species. Given this electron-hole and spin symmetry, and to avoid redundancy, we restrict our analysis to spin-up states throughout the discussion. Such control over gap closings implies a tunable transition between insulating and semimetallic regimes, offering a promising route to topological phase engineering. 
		
		Figure~\ref{f2} presents the spin and orbital Edelstein susceptibilities for the $K$ valley ($\zeta = +1$) as functions of the normalized mass asymmetry parameter \(\Delta_d / \Delta_{\text{so}}\), evaluated at three representative chemical potentials: \(\mu = 0\), \(\mu = \Delta_{+1,+1}\), and \(\mu > \Delta_{+1,+1}\), corresponding respectively to charge neutrality, the onset of conduction in one spin-valley sector, and finite electron doping. At charge neutrality \(\mu = 0\), in left panels, both the intraband spin and orbital Edelstein susceptibilities vanish, as expected due to particle-hole symmetry and the absence of net Fermi surface contributions. However, the interband spin susceptibility remains finite and anisotropic, highlighting the role of virtual interband transitions in generating spin responses even when the Fermi level lies in the gap. At the critical point \(\Delta_d / \Delta_{\text{so}} = -1\), the intraband responses show no significant features, whereas the interband components of the spin susceptibility exhibits a clear discontinuity.
		
		When the chemical potential is set at the band edge, \(\mu = \Delta_{+1,+1}\), as shown in the middle panels of Fig.~\ref{f2}, small but finite intraband contributions appear, whereas the interband responses remain nearly unaffected. Notably, the intraband spin susceptibility displays a sharp reversal feature near \(\Delta_d / \Delta_{\text{so}} = -1\), where the mass term for one spin-valley sector vanishes and the Dirac point becomes gapless, leading to vanishing out-of-plane components. This treatment is echoed in the orbital Edelstein response, which exhibits a characteristic antisymmetric peak near this critical point and vanishes at the critical point, indicating enhanced Berry curvature dipole contributions. It is important to emphasize that the orbital contribution to the intraband response is approximately three orders of magnitude larger than its spin counterpart.
		
		For \(\mu > \Delta_{+1,+1}\) (right panels of Fig.~\ref{f2}), the intraband contributions become more pronounced. The spin response develops step-like features across \(\Delta_d / \Delta_{\text{so}} = -1\) with a slight shift due to the additional energy introduced by the chemical potential, while the orbital susceptibility shows a clear sign change and saturation beyond the critical point. In contrast, the interband spin contribution evolves smoothly and remains largely unaffected by further doping, underscoring its robustness against changes in the Fermi level.\begin{figure}[t]
			\centering
			\includegraphics[width=0.9\linewidth]{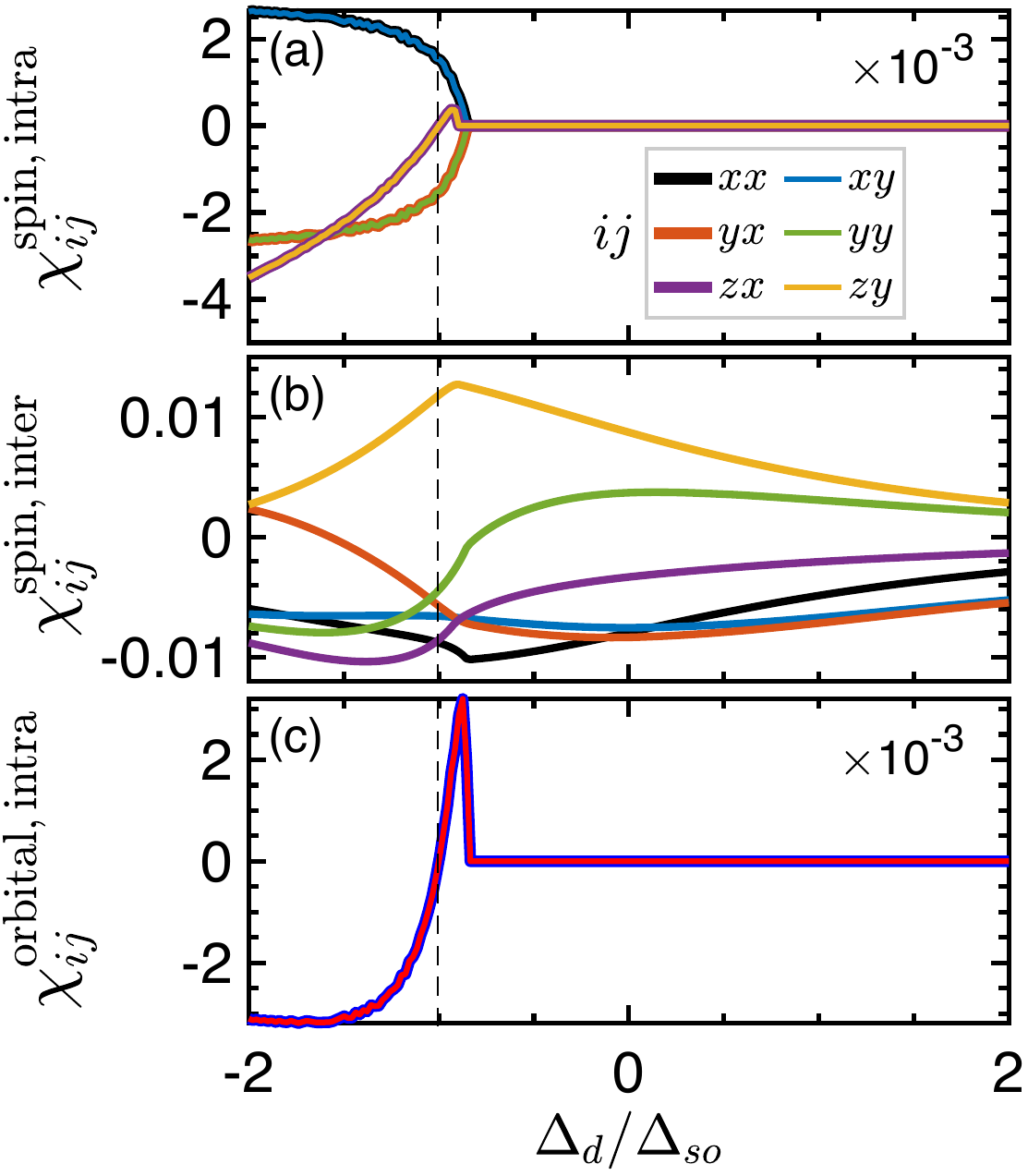}
			\caption{Edelstein response functions~(in arbitrary units) in monolayer jacutingaite Pt\(_2\)HgSe\(_3\) for the \( K' \) valley (\( \zeta = -1 \)) and spin-up $s = +1$ as a function of the dimensionless photoinduced mass term \( \Delta_d / \Delta_{\text{so}} \), computed for $\mu > \Delta_{-1,+1}$.  (a) Intraband spin Edelstein conductivity \( \chi^{\text{spin, intra}}_{ij} \). The vanishing trend around \( \Delta_d \sim -\Delta_{\text{so}} \) reflects the empty states below the chemical potential. (b) Interband spin Edelstein conductivity \( \chi^{\text{spin, inter}}_{ij} \), showing enhanced anisotropy and the discontinuity around the chemical potential. These interband contributions are especially sensitive to the topological character of the bands. (c) Intraband orbital Edelstein response \( \chi^{\text{orbital, intra}}_{ij} \), presenting a smooth crossover and sign change, indicating a buildup of nonequilibrium orbital magnetization induced by the helical drive.}
			\label{f3}
		\end{figure}
		
		Overall, these results demonstrate that the Edelstein response in Dirac materials with spin-orbit coupling and sublattice asymmetry is highly tunable through both mass inversion and chemical potential, with enhanced effects appearing near topological transitions where a mass term vanishes. The transition from a quantum spin Hall insulator to a semimetallic phase around the valley $K$ can be identified by a discontinuity in the spin Edelstein effect and a vanishing orbital Edelstein response. Altogether, these results demonstrate that both spin and orbital Edelstein effects provide powerful and experimentally accessible probes of light-tunable topological phase transitions in 2D Dirac materials. While alternative approaches such as RKKY-based methods—recently proposed in Ref.~\cite{PhysRevB.111.014440}—offer insightful routes to probing topological phase transitions in monolayer jacutingaite Pt\(_2\)HgSe\(_3\), our method provides a more direct and accessible means. This is because the Edelstein response stems locally from symmetry breaking and spin-orbit coupling, in contrast to the RKKY interaction, which depends on non-local spin-spin correlations mediated by conduction electrons. On the other hand, RKKY interactions typically require localized magnetic impurities, depend on long-range coherence, and are less sensitive to small gap closings. Their detection often demands indirect techniques such as scanning probe magnetometry or neutron scattering, whereas Edelstein effects can be probed via transport measurements.

		Figure~\ref{f3} demonstrates that the spin and orbital Edelstein response functions in valley $K'$ also serve as effective probes of the topological phase transition in monolayer jacutingaite Pt$_2$HgSe$_3$, particularly in the vicinity of the critical point $\Delta_d/\Delta_\text{so} \approx -1$, where the system becomes semi-metallic. For $\mu = 0$ and $\mu = \Delta_{-1,+1}$, similar features appear in the $K'$	valley, and we omit their detailed discussion here to avoid redundancy. However, for $\mu > \Delta_{-1,+1}$, in contrast to the valley $K$, where all intraband spin Edelstein components exhibit similar trends and responses, the behavior of $\chi_{ij}^{\text{spin, intra}}$ at the $K'$ valley displays a notable curvature reversal in the spin components $\sigma_x$ and $\sigma_y$. This behavior stems directly from the valley-selective nature of the light-induced potential $\zeta \Delta_d$ in Eq.~\eqref{eq_1}, as well as the asymmetry between the $x$ and $y$ components of the velocity operators in their coupling to the valley index, as defined in Eqs.~\eqref{eq_B1d} and~\eqref{eq_B1e}. Concurrently, the interband spin Edelstein response $\chi_{ij}^{\text{spin, inter}}$ develops increased anisotropy, indicative of the enhanced sensitivity of interband coherence to the underlying topological features of the Floquet-engineered bands.\begin{figure}[t]
			\centering
			\includegraphics[width=1\linewidth]{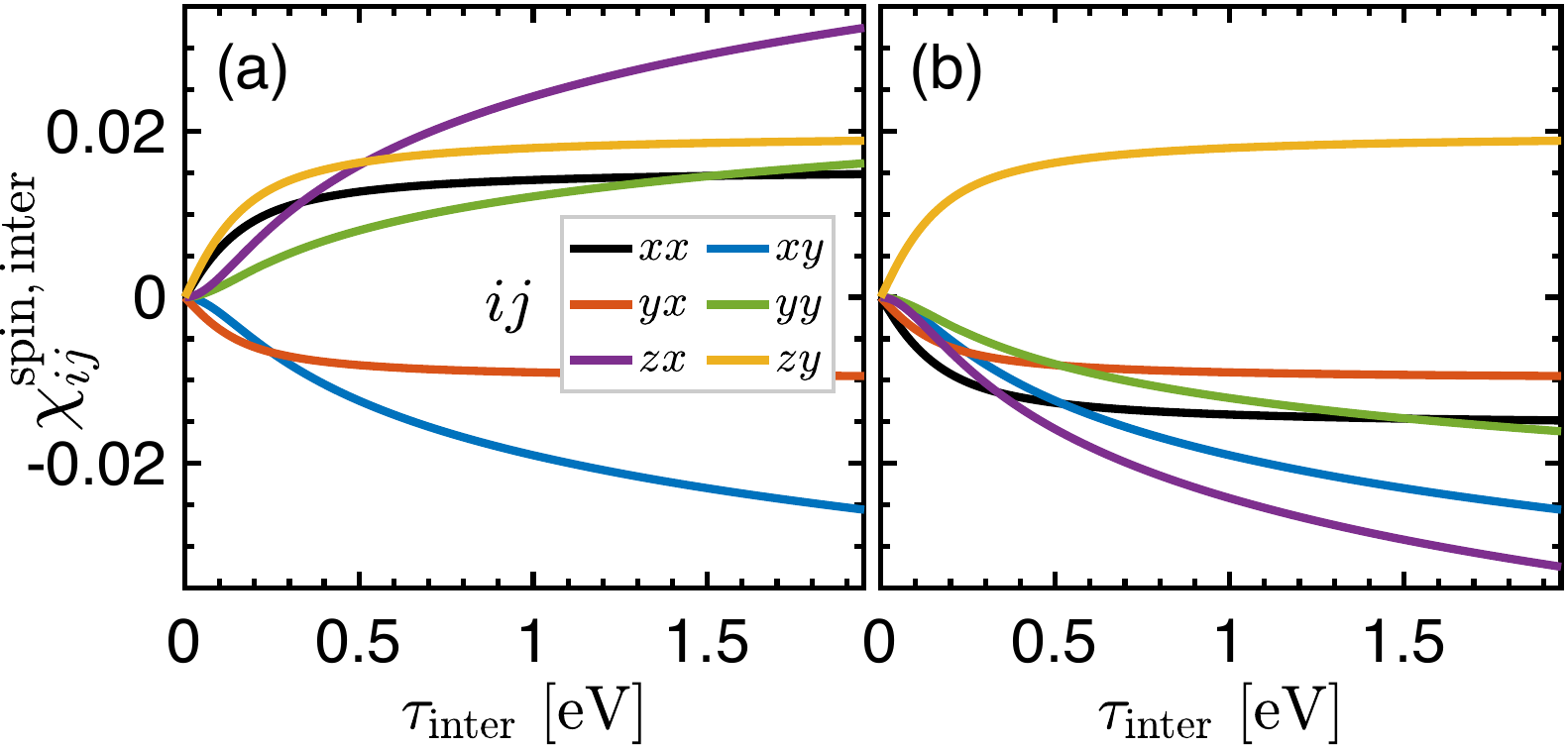}
			\caption{Interband spin Edelstein conductivity tensor \( \chi^{\text{spin, inter}}_{ij} \)~(in arbitrary units) in monolayer jacutingaite Pt\(_2\)HgSe\(_3\) as a function of the interband scattering time \( \tau_{\text{inter}} \), evaluated in units of eV. All panels correspond to a fixed light-induced potential \( \Delta_d = - \Delta_{\text{so}} \), while each subplot represents a distinct set of valley for spin-up \( s = +1\) for the photoinduced Dirac model: (a) $K$ and (b) $K'$.	Each plot shows the full tensor response \( ij = \{xx, xy, yx, yy\} \) and reveals the rich interplay between relaxation time and symmetry breaking. The non-monotonic and anisotropic trends highlight the sensitivity of the interband spin Edelstein effect to the scattering environment.}
			\label{f4}
		\end{figure}
		
		Finally, Fig.~\ref{f4} presents the interband spin Edelstein conductivity tensor, $\chi^{\text{spin, inter}}_{ij}$, in monolayer jacutingaite (Pt$_2$HgSe$_3$) as a function of the interband scattering time $\tau_{\text{inter}}$ at a fixed light-induced potential $\Delta_d = -\Delta_{\text{so}}$, where the phase transition occurs. In panel (a), associated with the $K$ valley, components such as $\chi_{xx}$ (black) and $\chi_{yy}$ (green) display a clear monotonic increase with $\tau_{\text{inter}}$, whereas the off-diagonal components demonstrate asymmetric trends. Specifically, $\chi_{xy}$ decreases strongly while $\chi_{yx}$ remains relatively stable. In contrast, the $K'$ valley shown in panel (b) exhibits a more gradual and saturating behavior for diagonal components, while the off-diagonal elements also show non-monotonic and anisotropic characteristics, albeit with reduced magnitude. The anisotropy observed between $xx$ and $yy$, as well as between off-diagonal components, is indicative of inversion symmetry breaking within the system. Furthermore, the valley-contrasting behavior underscores the sensitivity of the interband spin Edelstein effect to valley-specific dynamics, which is a hallmark of spin-valley coupled Dirac systems. The non-monotonic trends with respect to $\tau_{\text{inter}}$ highlight the intricate interplay between spin response and scattering environments. 
		
		\section{Summary}\label{s4}
		The optical modulation of topological phases in 2D quantum materials provides a promising platform for advancing spintronic and valleytronic technologies. In this work, we explored monolayer jacutingaite (Pt$_2$HgSe$_3$)—a robust quantum spin Hall insulator characterized by strong intrinsic spin-orbit coupling—subjected to off-resonant circularly polarized light. Through Floquet engineering, we demonstrated the emergence of light-induced topological transitions, namely the formation of spin- and valley-selective semimetallic states where the band gap vanishes selectively depending on spin and valley indices.
		
		To detect and characterize these transitions, we utilized the spin and orbital Edelstein effects, which capture the generation of non-equilibrium spin and orbital polarizations in response to an external electric field in non-centrosymmetric systems. Crucially, our approach does not rely on computing topological invariants such as Chern numbers; instead, we identify clear and universal signatures of topological transitions directly from transport observables. In particular, the transition from a gapped quantum spin Hall phase to a gapless semimetallic regime is marked by vanishing the orbital Edelstein susceptibility and a pronounced discontinuity in the spin Edelstein conductivity. Moreover, we examined the role of interband scattering processes by analyzing how the Edelstein responses evolve with varying scattering times across the phase boundaries. Our results reveal a consistent correlation between the relaxation dynamics and the critical behavior of spin susceptibility components near the topological transition.
		
		These findings highlight the Edelstein effect as a powerful and experimentally viable diagnostic tool for probing photoinduced topological phase transitions. They also provide a foundation for the use of non-equilibrium spin-orbit phenomena in mapping and controlling light-driven quantum phases in emerging 2D materials.
		
  	\section*{Acknowledgments}
		This research is funded by Vietnam National Foundation for Science and Technology Development (NAFOSTED) under grant number 103.01-2021.68. We are grateful to Mohsen Yarmohammadi for helpful discussions.
		\section*{Conflicts of interest}
			There are no conflicts to declare.
		\section*{Data Availability Statement}	
All data that support the findings of this study are included within the article.
		\appendix
		\section{Light-driven potential in monolayer Pt$_2$HgSe$_3$}\label{ap1}
		
		In this Appendix, we investigate the influence of helically polarized electromagnetic waves on the dynamics of Dirac fermions under a time-dependent electric field. The interaction between the system and the field is captured by a time-dependent vector potential, given by
		\begin{align}
			\vec{A}(t) = A_0 \left[\sin(\omega_d t), \cos(\omega_d t)\right],
		\end{align}where \( A_0 = \frac{E_0}{\sqrt{2} \omega_d} \) represents the amplitude of the vector potential, with \( E_0 \) being the electric field amplitude and \( \omega_d \) the frequency of the incident radiation. This vector potential modifies the system's Hamiltonian through minimal coupling, inducing periodic transitions between the system's eigenstates. Since the Hamiltonian is time-periodic, satisfying \( H_{\zeta,s}(t) = H_{\zeta,s}(t + T) \) with \( T = 2\pi / \omega_d \), we apply Floquet theory to describe the system's response \cite{GRIFONI1998229,PLATERO20041}.
		
		In the high-frequency regime, where the amplitude of the field remains small compared to the driving frequency, a perturbative expansion of the Floquet Hamiltonian leads to a simplified effective model. In this regime, interband transitions are suppressed, and the first-order correction to the Hamiltonian is expressed as~\cite{PhysRevB.84.235108,PhysRevB.82.235114,PhysRevB.84.235108,PhysRevLett.110.026603}
		\begin{align}
			H_{\rm F, \zeta, s} \approx H_0 + \frac{[H_{-1}, H_{+1}]}{\omega_d},
		\end{align}where the harmonics are defined by
		\begin{subequations}
			\begin{align}
				H_{0} &= \frac{1}{T} \int_0^T H^0_{\zeta,s}(\vec{k}, t) \, dt, \\
				H_{\pm 1} &= \frac{1}{T} \int_0^T H^0_{\zeta,s}(\vec{k}, t) e^{\pm i \omega_d t} \, dt.
			\end{align}
		\end{subequations}
		This correction induces a light-driven gap at the Dirac points, which modifies the system's electronic dispersion and affects transport properties. The induced gap is given by
		\begin{align}
			\Delta_d = \frac{e^2 A_0^2 v_{\rm F}^2}{\omega_d}\, .
		\end{align}
		
		\section{Spin and orbital Edelstein components in monolayer Pt$_2$HgSe$_3$}\label{ap2}
		In this Appendix, we provide all the components necessary for computing the spin and orbital Edelstein susceptibilities, expressed as
			\begin{subequations}
				\begin{align}
					O^{x,\,{\rm spin}}_{\pm \pm,\vec k} = {} &\pm \frac{\hbar}{2} \sin(\theta_{\vec k}) \cos(\phi_{\vec k})\, ,\\
					O^{y,\,{\rm spin}}_{\pm \pm,\vec k} = {} &\pm \frac{\hbar}{2} \sin(\theta_{\vec k}) \sin(\phi_{\vec k})\, ,\\
					O^{z,\,{\rm spin}}_{\pm \pm,\vec k} = {} &\pm \frac{\hbar}{2} \cos(\theta_{\vec k}) \, ,\\
					p^{x}_{\pm \pm,\vec k} = {} &\pm \frac{m_{\rm e}\zeta v_{\rm F}}{\hbar} \sin(\theta_{\vec k}) \cos(\phi_{\vec k})\, ,\label{eq_B1d}\\
					p^{y}_{\pm \pm,\vec k} = {} &\pm \frac{m_{\rm e} v_{\rm F}}{\hbar} \sin(\theta_{\vec k}) \sin(\phi_{\vec k})\, ,\label{eq_B1e}\\
					p^{z}_{\pm \pm,\vec k} = {} &0\, ,\\
					O^{x,\,{\rm spin}}_{\pm \mp,\vec k} = {} &\frac{\hbar}{2}\left(\sin^2(\theta_{\vec k}/2)e^{\mp i\phi_{\vec k}}-\cos^2(\theta_{\vec k}/2)e^{\pm i\phi_{\vec k}}\right)\, ,\\
					O^{y,\,{\rm spin}}_{\pm \mp,\vec k} = {} &\frac{\pm i\hbar}{2}\left(\sin^2(\theta_{\vec k}/2)e^{\mp i\phi_{\vec k}}+\cos^2(\theta_{\vec k}/2)e^{\pm i\phi_{\vec k}}\right)\, ,\\
					O^{z,\,{\rm spin}}_{\pm \mp,\vec k} = {} &\frac{\hbar}{2}\sin(\theta_{\vec k})\, ,\\
					p^{x}_{\pm \mp,\vec k} = {} &\frac{\zeta m_{\rm e} v_{\rm F}}{\hbar}\left(\sin^2(\theta_{\vec k}/2)e^{\mp i\phi_{\vec k}}-\cos^2(\theta_{\vec k}/2)e^{\pm i\phi_{\vec k}}\right)\, ,\\
					p^{y}_{\pm \mp,\vec k} = {} &\frac{\pm i m_{\rm e}v_{\rm F}}{\hbar}\left(\sin^2(\theta_{\vec k}/2)e^{\mp i\phi_{\vec k}}-\cos^2(\theta_{\vec k}/2)e^{\pm i\phi_{\vec k}}\right),\\
					p^{z}_{\pm \mp,\vec k} = {} &0\, .
				\end{align}
		\end{subequations}
	}
	\bibliography{bib_PRB.bib}
	
\end{document}